# High-performance InSe Transistors with Ohmic Contact Enabled by Nonrectifying-barrier-type Indium Electrodes

Yu-Ting Huang[1,2], Yi-Hsun Chen[1], Yi-Ju Ho[1], Shih-Wei Huang[1], Yih-Ren Chang[3], Kenji Watanabe[4], Takashi Taniguchi[4], Hsiang-Chih Chiu[5], Chi-Te Liang[2], Raman Sankar[6], Fang-Cheng Chou[6], Chun-Wei Chen[3]*, and Wei-Hua Wang[1]*

[1]Institute of Atomic and Molecular Sciences, Academia Sinica, Taipei 106, Taiwan

[2]Department of Physics, National Taiwan University, Taipei 106, Taiwan

[3]Department of Materials Science and Engineering, National Taiwan University, Taipei 106, Taiwan

[4]National Institute for Materials Science, 1-1 Namiki, Tsukuba, Ibaraki 305-0044, Japan

[5]Department of Physics, National Normal Taiwan University, Taipei 106, Taiwan

[6]Center of Condensed Matter Sciences, National Taiwan University, Taipei 106, Taiwan

ABSTRACT

The electrical contact to two-dimensional (2D)-semiconductor materials are decisive to the electronic performance of 2D-semiconductor field-effect devices (FEDs). The presence of a Schottky barrier often leads to a large contact resistance, which seriously limits the channel conductance and carrier mobility measured in a two-terminal geometry. In contrast, ohmic contact is desirable and can be achieved by the presence of a nonrectifying or tunneling barrier. Here, we demonstrate that an nonrectifying barrier can be realized by contacting indium (In), a low work function metal, with layered InSe because of a favorable band alignment at the In–InSe interface. The nonrectifying barrier is manifested by ohmic contact behavior at $T = 2$ K and a low barrier height, $\Phi_B = 50$ meV. This ohmic contact enables demonstration of an ON-current as large as 410 µA/µm, which is among the highest values achieved in FEDs based on layered semiconductors. A high electron mobility of 3,700 and 1,000 cm$^2$/V·s is achieved with the two-terminal In–InSe FEDs at $T = 2$ K and room temperature, respectively, which can be attributed to enhanced quality of both conduction channel and the contacts. The improvement in the contact quality is further proven by an X-ray photoelectron spectroscopy study, which suggests that a reduction effect occurs at the In–InSe interface. The demonstration of high-performance In–InSe FEDs indicates a viable interface engineering method for next-generation, 2D-semiconductor-based electronics.

KEYWORDS

2D semiconductors, high carrier mobility, ohmic contact, nonrectifying barrier, nanoelectronics

# INTRODUCTION

2D-semiconductor materials have the potential to influence nanoelectronics because of their novel functionalities.[1] Owing to their atomically thin layer and excellent gate coupling, 2D-semiconductor materials are attractive because of their lack of short-channel effects,[2-4] which considerably lower device performances. Moreover, the flexibility and transparency of 2D-semiconductor materials offer opportunities for the development of next-generation electronics.[5] In this study, we investigate the transport properties of layered InSe, a 2D semiconductor composed of metal chalcogenide, which exhibits intriguing properties, including a high carrier mobility in bulk[6] and in multilayer forms,[7-9] promising optoelectronic characteristics,[10-12] and quantum transport behaviors.[7]

To realize an ohmic contact by reducing the contact resistance between a semiconducting channel and metal is one of the most important issues in enhancing the performance of 2D-semiconductor-based field-effect devices (FEDs).[13-20] The presence of a Schottky barrier, which forms at the metal/semiconductor interface, is common in 2D-semiconductor FEDs and can cause a large contact resistance. Moreover, InSe is chemically unstable under ambient conditions,[9] which results in considerable degradation of its contact properties due to numerous surface defects and a large contact barrier. As a result, the large contact resistance caused by the Schottky barrier and/or material degradation at an interface seriously limits the channel conductance and carrier mobility measured in a two-terminal geometry. Numerous techniques, including multi-probe measurements,[7-8] tunneling contact,[21] and fabrication in an inert atmosphere,[7] have been developed to address this contact issue.

An ohmic contact is desirable for FEDs and can be achieved with nonrectifying barriers or a tunneling barrier. However, a feasible method to achieve good contact in a layered semiconductor FEDs with a nonrectifying barrier is rarely discussed. Here, we demonstrate that an nonrectifying-barrier-type ohmic contact can be realized in a layered semiconductor FEDs by contacting indium (In) with layered InSe because of a favorable band alignment at the In–InSe interface. This nonrectifying barrier enables the realization of ohmic contact with low-temperature, linear current-voltage behavior from the subthreshold to the high carrier density regime with a small barrier height. The optimized ohmic contact allows us to achieve remarkable transport characteristics, including a high electron mobility of $\mu = 3,700$ cm$^2$/V·s and a large ON-current of 410 µA/µm, which were both measured in a two-terminal geometry. Notably, the high-performance InSe FEDs are realized by a viable fabrication method under ambient conditions.

EXPERIMENTAL SECTION

The In–InSe FEDs were fabricated on hexagonal boron nitride (h-BN) flakes, which acted as substrates to reduce the extrinsic scattering sources at the bottom interface of the InSe.[22] These h-BN flakes were mechanically exfoliated onto a heavily n-doped Si wafers with 300-nm-thick SiO$_2$ layer for the subsequent InSe transfer. Few-layer InSe was first exfoliated onto polydimethylsiloxane (PDMS), and then, the targeted InSe flakes (~10-15 nm thickness) were identified using the contrast differences in the optical microscopy (OM) and atomic force microscope images and Raman spectroscopy results (Supporting information S1). To achieve staking structures while avoiding chemical residues that may result from a wet-transfer technique, we adopted a dry-transfer technique[23] to transfer the InSe onto the h-BN flakes. We annealed the InSe samples in a furnace at 300 °C with Ar 95% and H$_2$ 5% forming gas to remove the PDMS residues. Finally, we used a shadow mask method to deposit the contact metal to avoid possible contamination caused by a resist residue. The In contact was deposited in two steps: first, the samples were annealed at 90 °C in a vacuum of $1 \times 10^{-8}$ Torr followed by the deposition of a 3-nm In layer over the whole InSe flakes as an interfacial In layer at a base pressure of $1 \times 10^{-7}$ Torr; second, we align the shadow mask with the InSe flakes and In/Au (15 nm/40 nm) was subsequently deposited on the contact region at a base pressure of $1 \times 10^{-7}$ Torr. Between the two metal deposition processes, the samples were exposed to the air for the alignment of the shadow mask and the exposure time was within 30 min. For the control samples without the 3 nm In layer, only In/Au (15 nm/40 nm) was deposited for metallic contacts. A schematic of the contact regime of the device structure and an OM image of a fabricated InSe FED are shown in Figure 1a and the inset of Figure 2a, respectively.

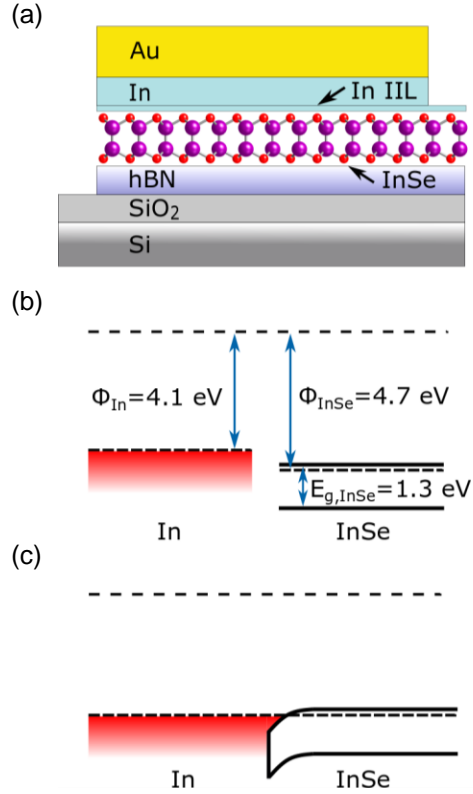

**Figure 1.** (a) A schematic of the device structure at the contact regime. The layered InSe material is interfaced with an In layer on top to create a nonrectifying barrier at the contact regime. Energy band diagrams of the In–InSe interface before (b) and after (c) the two materials are in contact are depicted in the schematics. When In and InSe are in contact, electron transfer causes a downward bending of the InSe CB band edge near its interfaces with In. This favorable band alignment results in an ideal nonrectifying barrier, and electrons can easily flow across the junction under both positive and negative bias voltages.

The nonrectifying barrier was realized by choosing a low work function metal, In, as the contact metal for n-type InSe because of the favorable band alignment predicted at the In–InSe interface. As shown in Figure 1b, In exhibits a work function of 4.12 eV,[24] which is lower than the work function of InSe, i.e., 4.7 eV.[25] The InSe thickness used in this study was approximately 10-15 nm, and the bandgap of InSe was the same as that of bulk InSe (1.3 eV) when a quantum confinement effect is absent.[26-27] When In and InSe are in contact, electrons from the In contact layer flow into InSe to reach thermal equilibrium, causing the InSe CB band edge to bend downward (electron doping) near its interfaces with In, as depicted in Figure 1c. The band bending effectively makes the semiconductor surface exhibit more n-type properties, leading to a high surface charge density at the interface. This band alignment results in an nonrectifying barrier, and electrons can easily flow across the junction when both positive and negative voltages are applied to the metal contact with respect to the semiconductor, leading to an ohmic contact.[28] In this study, we show that the improvement in the contact and transport properties can only be realized by incorporating an In interfacial layer (IIL) in addition to a 15-nm In layer, even though a nonrectifying barrier is expected at the In–InSe interface based on the band alignment. In addition to the band alignment, the contact barrier can be further reduced through the metallization of InSe by substantial orbital overlapping[29-30] or strong In–In bonding at the interface.[31]

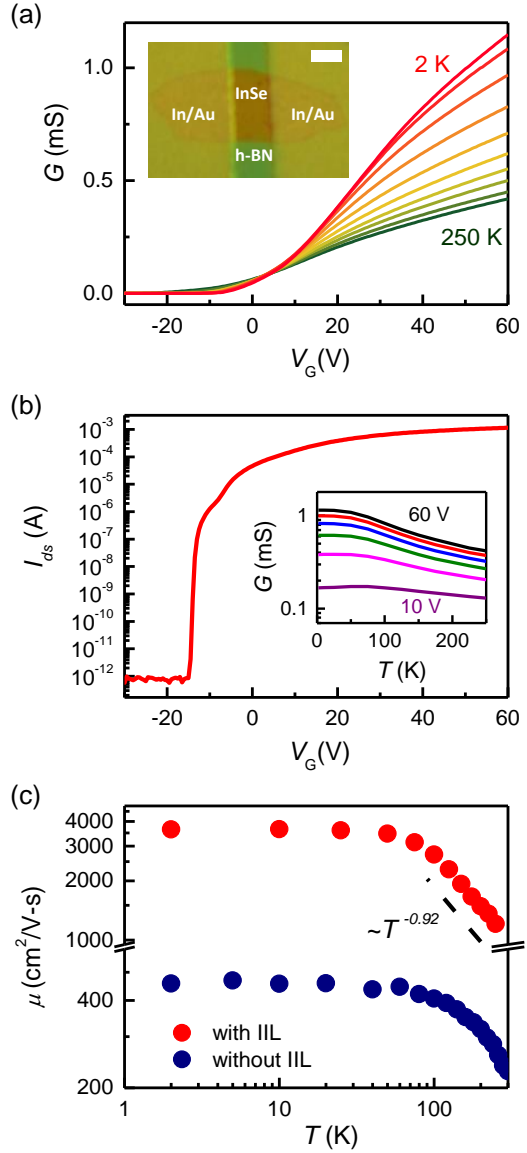

**Figure 2.** (a) The conductance as a function of the backgate voltage for an In–InSe FED with an In interfacial layer (IIL) (sample A) at $T$ values ranging from $T = 2$ K to $T = 250$ K, showing a noticeable MIT and dominant metallic behavior in the turn-on regime. $V_{ds} = 1$ V and the channel width and length of sample A are 8 and 13 μm, respectively. Inset: an OM image of a two-terminal In–InSe FED. The scale bar is 5 μm. (b) The logarithmic transfer curve for sample A, showing a large on/off ratio of $10^9$ at $T = 2$ K. Inset: the measured $G$ of sample A as a function of $T$ for $V_G$ values ranging from 0 V to 60 V, indicating metallic behavior. (c) A comparison of the field-effect

mobility as a function of $T$ for sample A (red circles) and sample B (blue circles). Sample A manifests a high $\mu_{imp} = 3{,}700$ cm²/V·s at a low $T$ in a two-terminal geometry and a phonon-limited, $T$-dependent mobility. The channel width and length of sample B are 9 and 15 μm, respectively.

RESULTS AND DISCUSSION

We first demonstrate the enhanced transport characteristics due to the incorporation of an IIL. Figure 2a shows the conductance, $G$, as a function of $V_G$ for an In–InSe FED with an IIL (sample A) at different $T$ values. The drain–source voltage ($V_{ds}$) is 1 V and the channel width and length of sample A are 8 and 13 μm, respectively. Sample A manifests a noticeable metal-insulator transition (MIT) with metallic behavior at a high carrier density and insulating behavior at a low carrier density. The transition occurs at a small carrier density, $n = 4 \times 10^{10}$ cm⁻² ($V_{MIT} - V_{th} = 0.6$ V) and $n = 7 \times 10^{11}$ cm⁻² ($V_{MIT} - V_{th} = 11$ V) for $T = 2$ K and $T = 250$ K, respectively, indicating a very low level of disorder in the InSe FED.[21] As a result, the sample shows a metallic $T$ dependence behavior in the majority of the turn-on regime. The transition point occurs at G = 83 μS ≈ $2e^2/h$, which can be attributed to either a critical quantum phenomenon or a classical percolation transition.[32-34] As a comparison, a control InSe FED without an IIL (sample B) shows a gradually shifting crossover point with a $T$ dependence without a well-defined MIT and insulating behavior over a wide $V_G$ range (Supporting information S2). Figure 2b shows the logarithmic transfer curve, $\log(I_{ds})$ as a function of $V_G$, for sample A at $T = 2$ K. The sample exhibits a typical turn-on behavior with a large on/off ratio of $10^9$ and the subthreshold swing of 230 mV/dec at $T = 2$ K. At $T = 250$ K, the on/off ratio and the subthreshold swing of sample A are $10^7$ and 5.6 V/dec, respectively (Supporting information S2).

Figure 2c compares the field-effect mobility, $\mu_{FE}$, as a function of $T$ for the InSe FED with (sample A) and without (sample B) an IIL. The $\mu_{FE}$ increases with the decreasing $T$ and saturates at a constant value, $\mu_{imp}$, for $T < 20$ K. At a low $T$, the phonon scattering effect is negligible, and the carrier scattering sources are mainly impurities, including long-range Coulomb impurities and short-range atomic defects.[35-36] We observe a remarkably high $\mu_{imp} = 3,700$ cm$^2$/V·s for sample A in a two-terminal geometry, indicating very minimal carrier scattering due to charged impurities and disorder in the In–InSe FED. Moreover, considering that $G$ is measured in a two-terminal configuration where contact resistance is relevant, the observed high mobility suggests a high contact quality. We note that the two-terminal mobility in our InSe device is higher than the previously reported value,[8-9,37] which highlights the importance of employing the In–InSe contact scheme and an IIL to improve the contact. In comparison, sample B shows a low $\mu_{imp} = 450$ cm$^2$/V·s. At high $T$ values, $\mu_{FE}$ decreases with the increasing $T$ and is governed by a power law, $\mu_{ph} \propto T^{-\gamma}$, indicating a phonon-limited, $T$-dependent mobility. The exponent for sample A is $\gamma = 0.92$, which reasonably agrees with the acoustic phonon scattering in InSe, i.e., $\gamma = 1$.[38] Sample A exhibits a high $\mu_{FE} = 1,000$ cm$^2$/V·s at room $T$, which favorably compares with the previously reported two-terminal mobility of few-layer InSe FEDs. The measured $G$ of sample A as a function of $T$ for $V_G$ values ranging from 0 V to 60 V is shown in the inset of Figure 2b. For $V_G \geq 10$ V, $G$ increases with the decreasing $T$, indicating a metallic $T$ dependence. We note that the metallic behavior arises at a small carrier density of $9 \times 10^{11}$ cm$^{-2}$, which is smaller than that of high-performance MoS$_2$ devices.[39]

The $T$-dependent output characteristics ($I_{ds}$–$V_{ds}$ curves) allow us to verify the decisive role of the IIL in the contact quality. Figure 3a shows the $I_{ds}$–$V_{ds}$ curves of the InSe FED with an IIL (sample A) at $T = 2$ K with $V_G$ values ranging from -10 V to 60 V. All these $I_{ds}$–$V_{ds}$ curves are linear and symmetric, indicating an ohmic contact. Notably, the ohmic behavior is realized under two conditions: (a) in the metallic and insulating regimes, i.e., $V_G < V_{MIT} = 4.5$ V, and (b) at a low $T$ where thermally activated electrons are absent, which both indicate a very small contact barrier. In comparison, sample B exhibits a non-linear response at $T = 2$ K, as shown in Figure 3b, suggesting a larger contact barrier.

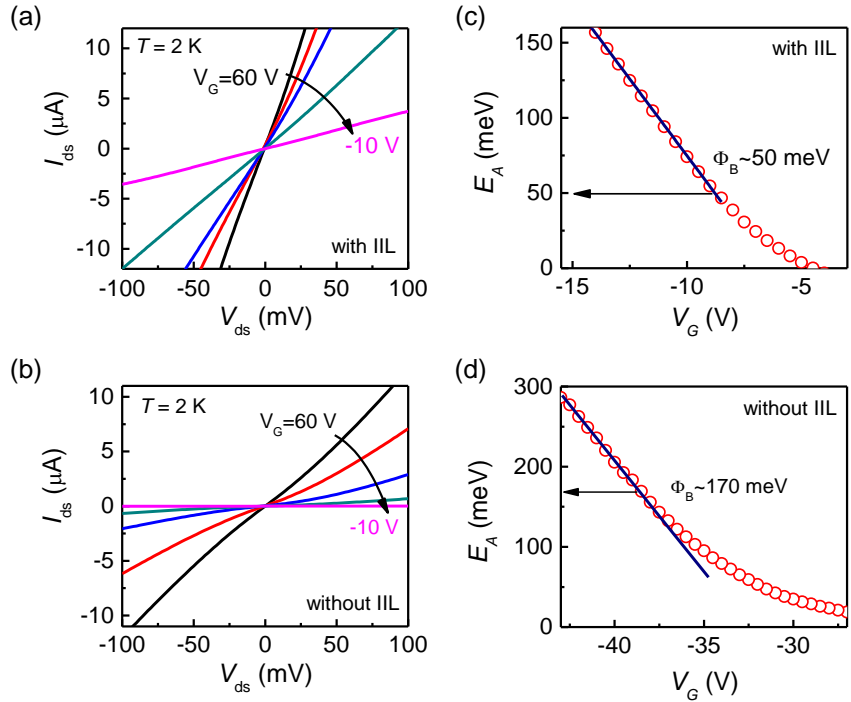

**Figure 3.** The $I_{ds}$–$V_{ds}$ curves of (a) sample A and (b) sample B at $T = 2$ K with $V_G$ values ranging from -10 V to 60 V. The $I_{ds}$–$V_{ds}$ curves of sample A are linear from the metallic regime into the insulating regime ($V_G < V_{MIT} = 4.5$ V) and at low $T$ values, indicating a very small contact barrier. In contrast, sample B exhibits a non-linear response at $T = 2$ K, suggesting a larger contact barrier.

The improvement in the contact by IIL is further confirmed by examining the effective contact barrier height. In a field-effect transistor (FET) with a Schottky barrier (SB) at both the source and drain, most of the voltage drop occurs at the reverse-biased contact, which basically determines the total drain current. Based on the thermionic emission model,[13] the drain current injected through a reverse-biased SB can be described as (Eq. 1)

$$I_d = A^* T^\alpha \exp\left[-\frac{E_A}{k_B T}\right]\left[1 - \exp(-\frac{qV_{ds}}{k_B T})\right] \quad (1)$$

where $A^*$ is the Richardson constant, $k_B$ is the Boltzmann constant, and α is an exponent of 3/2 for 2D semiconductors. By fitting the Arrhenius plot of $I_d/T^{3/2}$ (Supporting information S3), we can extract an activation energy, $E_A$, corresponding to the barrier the carriers must overcome. In the subthreshold regime, where thermionic emission dominates the drain current, the $E_A$ linearly depends on $V_G$. The flat band condition occurs when the band bending vanishes at a critical point, i.e., $V_G = V_{FB}$, at which the $E_A$ is equal to the barrier height, $\Phi_B$. At $V_G > V_{FB}$, tunneling starts to contribute to the drain current, causing $E_A$ to deviate from linearity. Figure 3c shows the $V_G$ dependence of $E_A$ for sample A with an IIL, yielding a barrier height of $\Phi_B = 50$ meV, which is comparable to the previously reported low barrier height of InSe.[9] In comparison, sample B exhibits a larger barrier height, $\Phi_B = 170$ meV, as shown in Figure 3d. We note that because the calculation of mobility in two-terminal geometry is related to the value of contact resistance, the small $\Phi_B$ observed in samples with an IIL can lead to the extracted high mobility aforementioned. To further verify the role of the IIL at the contact region, we fabricated another control sample with IIL deposited only on the InSe channel and not in the contact regime (Supporting information S4). The lack of enhanced transport properties in this control sample suggests the importance of introducing the IIL into the contact regime.

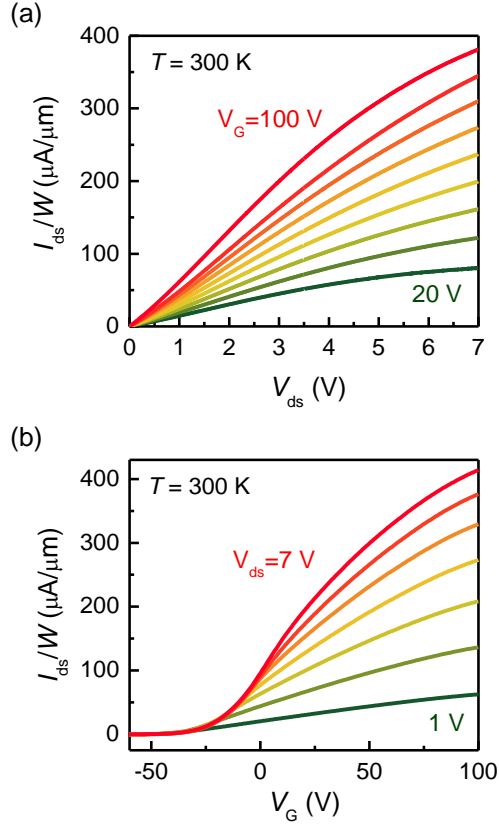

**Figure 4.** The transport characteristics of the In–InSe FEDs in a high electric field. (a) The output characteristics of an In–InSe FED (sample C) at room $T$ and different $V_G$ values. The channel width and length of sample C are 4 and 7 μm, respectively. The current saturation at a high bias voltage is observed with the highest ON-current of 380 μA/μm at $V_{ds} = 7$ V and $V_G = 100$ V. (b) The transfer characteristics of sample C at room $T$ and at different $V_{ds}$ values. The ON-current reaches a high value of 410 μA/μm with $V_{ds} = 7$ V and $V_G = 100$ V.

Next, we present the transport characteristics of In–InSe FEDs in a high electric field to examine the effect of incorporating the In–InSe contact and IIL. Figure 4a shows the output characteristics of an InSe FED (sample C) at room $T$ and at different $V_G$ values, and a linear $I_{ds}$–$V_{ds}$ relationship at a low $V_{ds}$ is observed. At a high $V_{ds}$ value, sample C manifests a high ON-current of 380 μA/μm at $V_{ds} = 7$ V and $V_G = 100$ V. We observe a linear $V_G$-dependence of the $I_{ds}$ in the saturation region, indicating that the velocity saturation governs the current saturation behavior.[40] The current saturation in the In–InSe FED is not significant, suggesting that a large saturation velocity is realized in the InSe sample. In contrast, the InSe FED without an IIL lacks the current saturation behavior and shows non-linear $I_{ds}$–$V_{ds}$ curves at low $V_{ds}$ values (Supporting information S5). We further show the transfer characteristics of sample C at room $T$ and at different $V_{ds}$ values in Figure 4b. The ON-current reaches a high value of 410 μA/μm at $V_G = 100$ V with $V_{ds} = 7$ V, which favorably compares to the previously reported values for 2D-semiconductor-based FEDs.[9, 40]

We further analyze the output characteristics to extract the saturation velocity, $v_{sat}$, by a simple semiempirical model, in which the $I_{ds}$ can be expressed as (Eq. 2)

$$I_{ds} = \frac{W}{L}\mu C_{ox}(V_G - V_{th})\frac{V_{ds}}{[1+(V_{ds}\mu/Lv_{sat})^\alpha]^{1/\alpha}} \qquad (2)$$

where $C_{ox}$ is the gate capacitance. The output characteristics are reasonably described by the equation, yielding $v_{sat} = 1.2 \times 10^7$ cm/s for $V_G = 80$ V (Supporting information S6), and the critical electric field, $\xi_{cr} = v_{sat}/\mu$, can be estimated as $2.4 \times 10^4$ V/cm. Because this expression is derived without considering the contact resistance, the good fit of the data suggests that the contact resistance is insignificant in the In–InSe FEDs. The large saturation velocity and current saturation in the In–InSe FEDs indicate the devices are enhanced due to two aspects of the In–InSe contact and IIL, the low contact resistance and the interface of the InSe channel exhibiting a low impurity density, which allow the realization of a high conductivity under high fields.[41] For a channel length of a few micrometers, the mobility can properly describe carrier transport in a low electric field. However, for a gate length on the order of 100 nm, FETs generally operate at a saturated carrier velocity under a high field, rendering the saturation velocity an important measure of device performance in this regime.[2] The extracted $v_{sat}$ value is comparable to that of Si and GaAs and higher than that of few-layer MoS$_2$,[40] showing the potential of employing InSe to operate in the short-channel regime with high drain current.

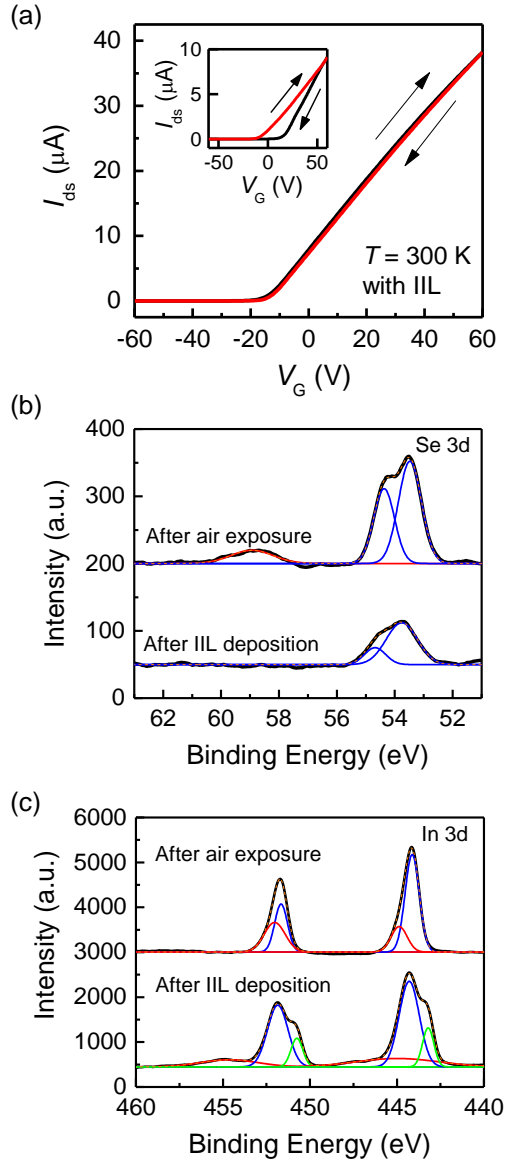

**Figure 5.** A comparison of the transfer characteristics of (a) an In–InSe FED with an In interfacial layer (IIL), which exhibits negligible hysteresis, and a control In–InSe FED without an IIL (inset). The applied drain–source voltage is 0.1 V. (b) The XPS data show a comparison of the Se 3d peaks before and after the deposition of an IIL. The Se 3d core level corresponding to Se–O species (~59 eV) is not detectable after the deposition of an IIL. (c) A comparison of the In 3d peaks before and after the deposition of an IIL. The two broad peaks in the high-energy region of the In $3d_{3/2}$ and In $3d_{5/2}$ core levels may be attributed to the formation of an In oxide complex.

To further demonstrate the effect of incorporating an IIL, we discuss the hysteresis effect of the InSe FEDs by showing the drain–source current ($I_{ds}$) as a function of the backgate voltage ($V_G$) (transfer characteristics) for a control InSe FED without an IIL in the inset of Figure 5a. The control InSe FED shows a large hysteresis between the upward and downward sweeping of $V_G$, which is commonly observed in samples with trap states at the interfaces.[42] In contrast, a very small hysteresis is observed in an InSe FED with an IIL, as shown in Figure 5a. The absence of hysteresis indicates that both the surface of the InSe channel and the InSe–h-BN interface exhibit negligible charge traps, which is favorable for carrier transport with low charge scattering.

Although a nonrectifying barrier at the In–InSe interface is expected based on the band alignment, the enhanced contact and transport properties are realized by incorporating an IIL. We performed an X-ray photoelectron spectroscopy (XPS) study of the InSe bulk crystals to investigate the role of the IIL in enhancing the transport characteristics of the InSe FEDs. XPS detects the chemical composition of a surface, allowing us to verify the effect due to oxidation and post-deposition of In. A bulk InSe sample was exposed to ambient conditions for 5 days to allow the reactant to be oxidized so a measurable XPS signal could be obtained. We observed the core levels of In 3d and Se 3d, and the blueshifts of the peaks indicate InSe oxidation caused by the doping effect (Supporting information S7). We then deposited a 3-nm In layer onto the bulk InSe employing the same procedures used for the InSe FED samples. Figure 5b compares the Se 3d peak before and after the deposition of the IIL. Notably, the Se 3d core level corresponding to Se–O species (~59 eV)[43-45] is no longer detectable after the deposition of the IIL, indicating the reduction of Se oxide. Figure 5c shows the In 3d peaks before and after the deposition of the IIL.

The emergence of the peaks in the high-energy region of the In 3d spectrum after air exposure can be attributed to surface oxidation (Supporting information S7).[46] After the deposition of the IIL, extra peaks at 450.8 and 443.2 eV are seen, indicating the presence of a metallic form of In.[47-48] Moreover, two broad peaks in the high-energy regions of the In $3d_{3/2}$ and In $3d_{5/2}$ core levels appear, suggesting the formation of an In oxide complex.[49-50] The important implication of the XPS data is that the post-deposited IIL can have a reduction effect on the oxidized Se on the surface. This reduction effect may improve the carrier transport in the contact region by reducing the charge scattering at the interface (Supporting information S8), resulting in the aforementioned, distinguished contact properties. On the other hand, the charge transport is affected by carrier scattering due to the presence of surface oxides in the samples without an IIL. Here, we note that the oxidation time for the InSe FEDs in the transport study is within 30 min, and much less oxidation is expected compared to that of the bulk InSe samples used for the XPS study. Therefore, complete reduction of the surface oxide is envisaged in the InSe FEDs used for the transport studies.

CONCLUSION

In conclusion, we have demonstrated ohmic contact for layered InSe with a nonrectifying barrier, resulting in high-quality In–InSe FEDs. The nonrectifying barrier at the InSe contact regime can be realized by employing In as the contact metal and an IIL as the interfacial layer owing to a favorable band alignment. This nonrectifying barrier consequently leads to the ohmic contact, as evidenced by the low-temperature linear current-voltage behavior and small barrier height, $\Phi_B = 50$ meV. With this optimized ohmic contact, we achieve distinguished transport characteristics for In–InSe FEDs in a two-terminal geometry, including a high electron mobility of $\mu_{FE} = 3,700$

cm$^2$/V·s and a high ON-current of 410 μA/μm. The presented high performance of In–InSe FEDs by a viable method highlights a unique opportunity to employ 2D-semiconductor materials for nanoelectronics applications.

**Supporting Information**.

Characterization of the InSe flakes. Metal-insulator transition. Estimation of the barrier height in the InSe FEDs. InSe FED with an IIL deposited only on the channel. $I_{ds}-V_{ds}$ curves of an InSe FED without IIL deposition. Estimation of the saturation velocity from the current saturation behavior. XPS study of InSe oxidation. Implications of the reduction effect and estimation of the transfer length. Chemical stability of the InSe samples studied by Raman spectroscopy.

**Corresponding Author**

*(W.-H. Wang) Tel: +886-2-2366-8208, Fax: +886-2-2362-0200, wwang@sinica.edu.tw; (C.-W. Chen) Tel: +886-2-3366-5205 Fax: +886-2-2363-4562, chunwei@ntu.edu.tw.

ACKNOWLEDGMENT

W. W. thanks Prof. Lain-Jong Li for insightful discussions. This work was supported by the Ministry of Science and Technology of Taiwan, R.O.C. under contract numbers MOST 103-2112-M-001-020-MY3 and MOST 106-2112-M-001-003-MY3.

# REFERENCES

(1) Wang, Q. H.; Kalantar-Zadeh, K.; Kis, A.; Coleman, J. N.; Strano, M. S. Electronics and Optoelectronics of Two-Dimensional Transition Metal Dichalcogenides. *Nat Nanotechnol* **2012,** *7* (11), 699-712.
(2) Schwierz, F. Graphene Transistors. *Nat Nanotechnol* **2010,** *5* (7), 487-496.
(3) Chhowalla, M.; Jena, D.; Zhang, H. Two-Dimensional Semiconductors for Transistors. *Nat Rev Mater* **2016,** *1* (11), 16052.
(4) Yoon, Y.; Ganapathi, K.; Salahuddin, S. How Good Can Monolayer Mos2 Transistors Be? *Nano Lett.* **2011,** *11* (9), 3768-3773.
(5) Akinwande, D.; Petrone, N.; Hone, J. Two-Dimensional Flexible Nanoelectronics. *Nat Commun* **2014,** *5*, 5678.
(6) Segura, A.; Pomer, F.; Cantarero, A.; Krause, W.; Chevy, A. Electron-Scattering Mechanisms in N-Type Indium Selenide. *Phys Rev B* **1984,** *29* (10), 5708-5717.
(7) Bandurin, D. A.; Tyurnina, A. V.; Yu, G. L.; Mishchenko, A.; Zolyomi, V.; Morozov, S. V.; Kumar, R. K.; Gorbachev, R. V.; Kudrynskyi, Z. R.; Pezzini, S.; Kovalyuk, Z. D.; Zeitler, U.; Novoselov, K. S.; Patane, A.; Eaves, L.; Grigorieva, I. V.; Fal'ko, V. I.; Geim, A. K.; Cao, Y. High Electron Mobility, Quantum Hall Effect and Anomalous Optical Response in Atomically Thin Inse. *Nat Nanotechnol* **2017,** *12* (3), 223-227.
(8) Sucharitakul, S.; Goble, N. J.; Kumar, U. R.; Sankar, R.; Bogorad, Z. A.; Chou, F. C.; Chen, Y. T.; Gao, X. P. A. Intrinsic Electron Mobility Exceeding 10(3) Cm(2)/(V S) in Multilayer Inse Fets. *Nano Lett.* **2015,** *15* (6), 3815-3819.
(9) Ho, P. H.; Chang, Y. R.; Chu, Y. C.; Li, M. K.; Tsai, C. A.; Wang, W. H.; Ho, C. H.; Chen, C. W.; Chiu, P. W. High-Mobility Inse Transistors: The Role of Surface Oxides. *Acs Nano* **2017,** *11* (7), 7362-7370.
(10) Lei, S. D.; Ge, L. H.; Najmaei, S.; George, A.; Kappera, R.; Lou, J.; Chhowalla, M.; Yamaguchi, H.; Gupta, G.; Vajtai, R.; Mohite, A. D.; Ajayan, P. M. Evolution of the Electronic Band Structure and Efficient Photo-Detection in Atomic Layers of Inse. *Acs Nano* **2014,** *8* (2), 1263-1272.
(11) Mudd, G. W.; Svatek, S. A.; Hague, L.; Makarovsky, O.; Kudrynskyi, Z. R.; Mellor, C. J.; Beton, P. H.; Eaves, L.; Novoselov, K. S.; Kovalyuk, Z. D.; Vdovin, E. E.; Marsden, A. J.; Wilson, N. R.; Patane, A. High Broad-Band Photoresponsivity of Mechanically Formed Inse-Graphene Van Der Waals Heterostructures. *Adv. Mater.* **2015,** *27* (25), 3760-3766.
(12) Tamalampudi, S. R.; Lu, Y. Y.; Kumar, U. R.; Sankar, R.; Liao, C. D.; Moorthy, B. K.; Cheng, C. H.; Chou, F. C.; Chen, Y. T. High Performance and Bendable Few-Layered Inse Photodetectors with Broad Spectral Response. *Nano Lett.* **2014,** *14* (5), 2800-2806.
(13) Allain, A.; Kang, J. H.; Banerjee, K.; Kis, A. Electrical Contacts to Two-Dimensional Semiconductors. *Nat Mater* **2015,** *14* (12), 1195-1205.
(14) Das, S.; Chen, H. Y.; Penumatcha, A. V.; Appenzeller, J. High Performance Multilayer Mos$_2$ Transistors with Scandium Contacts. *Nano Lett.* **2013,** *13* (1), 100-105.
(15) Lopez-Sanchez, O.; Lembke, D.; Kayci, M.; Radenovic, A.; Kis, A. Ultrasensitive Photodetectors Based on Monolayer Mos2. *Nat Nanotechnol* **2013,** *8* (7), 497-501.
(16) Kappera, R.; Voiry, D.; Yalcin, S. E.; Branch, B.; Gupta, G.; Mohite, A. D.; Chhowalla, M. Phase-Engineered Low-Resistance Contacts for Ultrathin Mos2 Transistors (Vol 13, Pg 1128, 2014). *Nat Mater* **2014,** *13* (12), 1128-1134.
(17) Popov, I.; Seifert, G.; Tomanek, D. Designing Electrical Contacts to Mos2 Monolayers: A

Computational Study. *Phys. Rev. Lett.* **2012,** *108* (15), 156802.
(18) Fang, H.; Chuang, S.; Chang, T. C.; Takei, K.; Takahashi, T.; Javey, A. High-Performance Single Layered Wse2 P-Fets with Chemically Doped Contacts. *Nano Lett.* **2012,** *12* (7), 3788-3792.
(19) Du, Y. C.; Yang, L. M.; Liu, H.; Ye, P. D. D. Contact Research Strategy for Emerging Molybdenum Disulfide and Other Two-Dimensional Field-Effect Transistors. *Apl Mater* **2014,** *2* (9), 092510.
(20) Liu, W.; Kang, J. H.; Sarkar, D.; Khatami, Y.; Jena, D.; Banerjee, K. Role of Metal Contacts in Designing High-Performance Monolayer N-Type Wse2 Field Effect Transistors. *Nano Lett.* **2013,** *13* (5), 1983-1990.
(21) Cui, X.; Shih, E. M.; Jauregui, L. A.; Chae, S. H.; Kim, Y. D.; Li, B. C.; Seo, D.; Pistunova, K.; Yin, J.; Park, J. H.; Choi, H. J.; Lee, Y. H.; Watanabe, K.; Taniguchi, T.; Kim, P.; Dean, C. R.; Hone, J. C. Low-Temperature Ohmic Contact to Monolayer Mos2 by Van Der Waals Bonded Co/H-Bn Electrodes. *Nano Lett.* **2017,** *17* (8), 4781-4786.
(22) Dean, C. R.; Young, A. F.; Meric, I.; Lee, C.; Wang, L.; Sorgenfrei, S.; Watanabe, K.; Taniguchi, T.; Kim, P.; Shepard, K. L.; Hone, J. Boron Nitride Substrates for High-Quality Graphene Electronics. *Nat Nanotechnol* **2010,** *5* (10), 722-726.
(23) Yang, R.; Zheng, X. Q.; Wang, Z. H.; Miller, C. J.; Feng, P. X. L. Multilayer Mos2 Transistors Enabled by a Facile Dry-Transfer Technique and Thermal Annealing. *J Vac Sci Technol B* **2014,** *32* (6), 061203.
(24) Michaelson, H. B. Work Function of Elements and Its Periodicity. *J. Appl. Phys.* **1977,** *48* (11), 4729-4733.
(25) Kudrynskyi, Z. R.; Bhuiyan, M. A.; Makarovsky, O.; Greener, J. D. G.; Vdovin, E. E.; Kovalyuk, Z. D.; Cao, Y.; Mishchenko, A.; Novoselov, K. S.; Beton, P. H.; Eaves, L.; Patane, A. Giant Quantum Hall Plateau in Graphene Coupled to an Inse Van Der Waals Crystal. *Phys. Rev. Lett.* **2017,** *119* (15), 157701.
(26) Madelung, O. *Semiconductors : Data Handbook*, 3rd ed.; Springer: Berlin ; New York, 2004; p 691 p.
(27) Mudd, G. W.; Svatek, S. A.; Ren, T.; Patane, A.; Makarovsky, O.; Eaves, L.; Beton, P. H.; Kovalyuk, Z. D.; Lashkarev, G. V.; Kudrynskyi, Z. R.; Dmitriev, A. I. Tuning the Bandgap of Exfoliated Inse Nanosheets by Quantum Confinement. *Adv. Mater.* **2013,** *25* (40), 5714-5718.
(28) Neamen, D. A. *Semiconductor Physics and Devices : Basic Principles*, 2nd ed.; Irwin: Chicago, 1997; p xx, 618 p.
(29) Kang, J. H.; Liu, W.; Banerjee, K. High-Performance Mos2 Transistors with Low-Resistance Molybdenum Contacts. *Appl. Phys. Lett.* **2014,** *104* (9), 093106.
(30) Jin, H.; Li, J. W.; Wan, L. H.; Dai, Y.; Wei, Y. D.; Guo, H. Ohmic Contact in Monolayer Inse-Metal Interface. *2d Mater* **2017,** *4* (2), 025116.
(31) Bailey, M. S.; DiSalvo, F. J. The Synthesis and Structure of Ca2inn, a Novel Ternary Indium Nitride. *J. Alloys Compd.* **2003,** *353* (1-2), 146-152.
(32) Licciardello, D. C.; Thouless, D. J. Conductivity and Mobility Edges for 2-Dimensional Disordered Systems. *J Phys C Solid State* **1975,** *8* (24), 4157-4170.
(33) Radisavljevic, B.; Kis, A. Mobility Engineering and a Metal-Insulator Transition in Monolayer Mos2. *Nat Mater* **2013,** *12* (9), 815-820.
(34) Das Sarma, S.; Hwang, E. H. Two-Dimensional Metal-Insulator Transition as a Strong Localization Induced Crossover Phenomenon. *Phys Rev B* **2014,** *89* (23), 235423-1-235423-20.
(35) Das Sarma, S.; Adam, S.; Hwang, E. H.; Rossi, E. Electronic Transport in Two-Dimensional

Graphene. *Reviews of Modern Physics* **2011,** *83* (2), 407-470.
(36) Chen, J. H.; Jang, C.; Xiao, S.; Ishigami, M.; Fuhrer, M. S. Intrinsic and Extrinsic Performance Limits of Graphene Devices on Sio2. *Nat Nanotechnol* **2008,** *3* (4), 206-209.
(37) Feng, W.; Zheng, W.; Cao, W. W.; Hu, P. A. Back Gated Multilayer Inse Transistors with Enhanced Carrier Mobilities Via the Suppression of Carrier Scattering from a Dielectric Interface. *Adv. Mater.* **2014,** *26* (38), 6587-6593.
(38) Sun, C.; Xiang, H.; Xu, B.; Xia, Y. D.; Yin, J.; Liu, Z. G. Ab Initio Study of Carrier Mobility of Few-Layer Inse. *Appl Phys Express* **2016,** *9* (3), 035203.
(39) Cui, X.; Lee, G. H.; Kim, Y. D.; Arefe, G.; Huang, P. Y.; Lee, C. H.; Chenet, D. A.; Zhang, X.; Wang, L.; Ye, F.; Pizzocchero, F.; Jessen, B. S.; Watanabe, K.; Taniguchi, T.; Muller, D. A.; Low, T.; Kim, P.; Hone, J. Multi-Terminal Transport Measurements of Mos2 Using a Van Der Waals Heterostructure Device Platform. *Nat Nanotechnol* **2015,** *10* (6), 534-540.
(40) Fiori, G.; Szafranek, B. N.; Iannaccone, G.; Neumaier, D. Velocity Saturation in Few-Layer Mos2 Transistor. *Appl. Phys. Lett.* **2013,** *103* (23), 233509.
(41) Lee, J.; Ha, T. J.; Parrish, K. N.; Chowdhury, S. F.; Tao, L.; Dodabalapur, A.; Akinwande, D. High-Performance Current Saturating Graphene Field-Effect Transistor with Hexagonal Boron Nitride Dielectric on Flexible Polymeric Substrates. *Ieee Electron Device Letters* **2013,** *34* (2), 172-174.
(42) Late, D. J.; Liu, B.; Matte, H. S. S. R.; Dravid, V. P.; Rao, C. N. R. Hysteresis in Single-Layer Mos2 Field Effect Transistors. *Acs Nano* **2012,** *6* (6), 5635-5641.
(43) Cui, Y. J.; Abouimrane, A.; Lu, J.; Bolin, T.; Ren, Y.; Weng, W.; Sun, C. J.; Maroni, V. A.; Heald, S. M.; Amine, K. (De)Lithiation Mechanism of Li/Sesx (X=0-7) Batteries Determined by in Situ Synchrotron X-Ray Diffraction and X-Ray Absorption Spectroscopy. *J. Am. Chem. Soc.* **2013,** *135* (21), 8047-8056.
(44) Yan, Y.; Liao, Z. M.; Yu, F.; Wu, H. C.; Jing, G. Y.; Yang, Z. C.; Zhao, Q.; Yu, D. P. Synthesis and Field Emission Properties of Topological Insulator Bi2se3 Nanoflake Arrays. *Nanotechnology* **2012,** *23* (30), 305704.
(45) Iwakuro, H.; Tatsuyama, C.; Ichimura, S. Xps and Aes Studies on the Oxidation of Layered Semiconductor Gase. *Jpn J Appl Phys 1* **1982,** *21* (1), 94-99.
(46) Ho, C. H.; Lin, C. H.; Wang, Y. P.; Chen, Y. C.; Chen, S. H.; Huang, Y. S. Surface Oxide Effect on Optical Sensing and Photoelectric Conversion of Alpha-In2se3 Hexagonal Microplates. *Acs Appl Mater Inter* **2013,** *5* (6), 2269-2277.
(47) Powell, C. J. Recommended Auger Parameters for 42 Elemental Solids. *J. Electron. Spectrosc. Relat. Phenom.* **2012,** *185* (1-2), 1-3.
(48) Procop, M. Xps Data for Sputter-Cleaned In0.53ga0.47as, Gaas, and Inas Surfaces. *J. Electron. Spectrosc. Relat. Phenom.* **1992,** *59* (2), R1-R10.
(49) Mcguire, G. E.; Schweitz.Gk; Carlson, T. A. Study of Core Electron Binding-Energies in Some Group Iiia, Vb, and Vib Compounds. *Inorg. Chem.* **1973,** *12* (10), 2450-2453.
(50) Liu, W. K.; Yuen, W. T.; Stradling, R. A. Preparation of Insb Substrates for Molecular-Beam Epitaxy. *J Vac Sci Technol B* **1995,** *13* (4), 1539-1545.

TOC

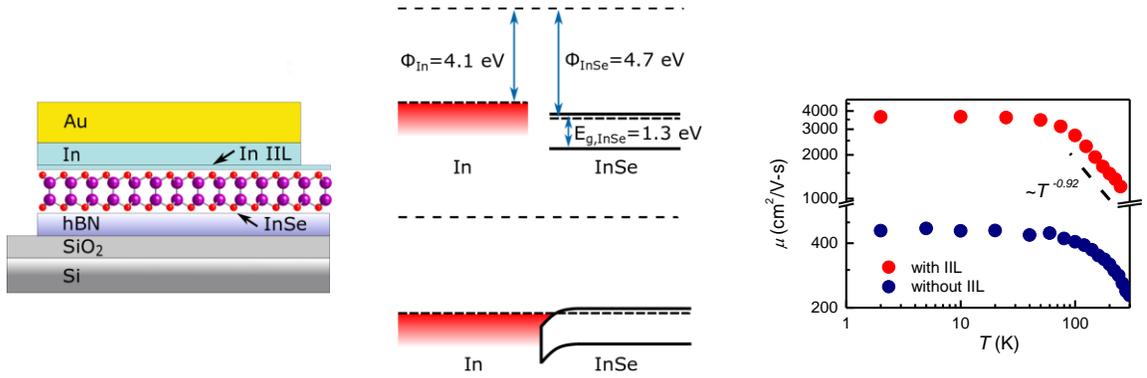

SUPPORTING INFORMATION

# High-performance InSe Transistors with Ohmic Contact Enabled by Nonrectifying-barrier-type Indium Electrodes

*Yu-Ting Huang[1,2], Yi-Hsun Chen[1], Yi-Ju Ho[1], Shih-Wei Huang[1], Yih-Ren Chang[3], Kenji Watanabe[4], Takashi Taniguchi[4], Hsiang-Chih Chiu[5], Chi-Te Liang[2], Raman Sankar[6], Fang-Cheng Chou[6], Chun-Wei Chen[3]\*, and Wei-Hua Wang[1]\**

[1]Institute of Atomic and Molecular Sciences, Academia Sinica, Taipei 106, Taiwan

[2]Department of Physics, National Taiwan University, Taipei 106, Taiwan

[3]Department of Materials Science and Engineering, National Taiwan University, Taipei 106, Taiwan

[4]National Institute for Materials Science, 1-1 Namiki, Tsukuba, Ibaraki 305-0044, Japan

[5]Department of Physics, National Normal Taiwan University, Taipei 106, Taiwan

[6]Center of Condensed Matter Sciences, National Taiwan University, Taipei 106, Taiwan

Corresponding Author
*(W.-H. Wang) wwang@sinica.edu.tw; (C.-W. Chen) chunwei@ntu.edu.tw.

## S1. Characterization of the InSe flakes

The InSe crystals used were grown by the Bridgman method. The InSe flakes were characterized by AFM and Raman spectroscopy. The height profile of an InSe/h-BN staking structure obtained by AFM is shown in Figure S1a. The thicknesses of the InSe and h-BN flakes are 10 nm and 50 nm, respectively. Figure S1b shows the Raman spectra of an InSe flake (10 nm). The InSe flake exhibits four characteristic peaks at 115, 178, 199, and 226 cm$^{-1}$, which are attributed to the Raman modes of $A_1'(\Gamma_1^2)$, $E'(\Gamma_3^1)$- TO, $A_2''(\Gamma_1^1)$- LO, and $A_1'(\Gamma_1^3)$, respectively.[1,2,3] The presence of these characteristic Raman peaks indicates the highly crystalline structure of the InSe flakes.

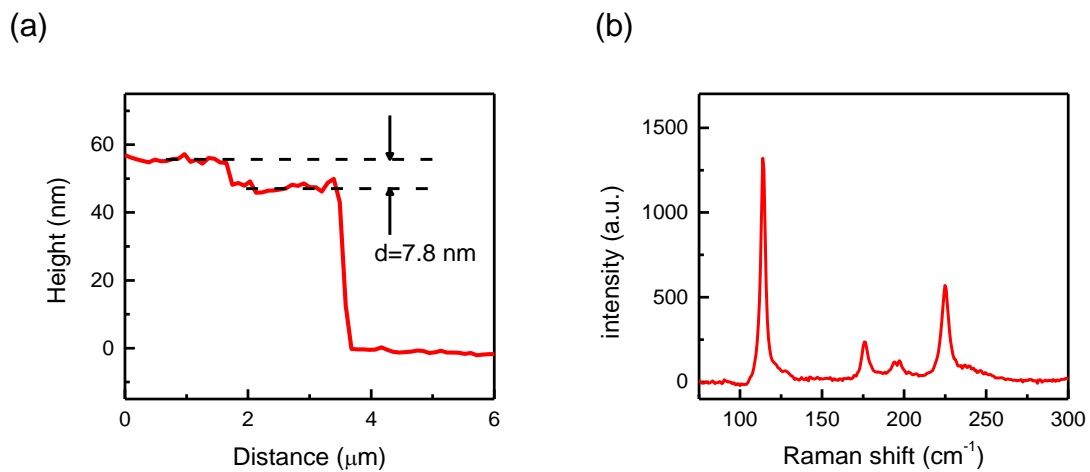

**Figure S1.** (a) Height profile an InSe/h-BN staking structure. (b) Raman spectra of an InSe flake with the characteristic peaks.

## S2. Metal-insulator transition and transfer characteristics

Figure S2a shows the $T$-dependent transfer characteristics of InSe FEDs (sample D) with an In interfacial layer (IIL), which exhibit a clear metal-insulator transition (MIT) phenomenon at $V_G = -10$ V. In comparison, an InSe FED without an IIL (Sample B in the main text) shows a gradually shifting crossover point in the $T$ dependence without a well-defined MIT, as shown in Figure S2b. This complicated behavior may be due to a sizable contact resistance with a $T$-dependent barrier height. Moreover, the insulating behavior of the control InSe FED exhibits a wider $V_G$ range, suggesting a lower device quality.

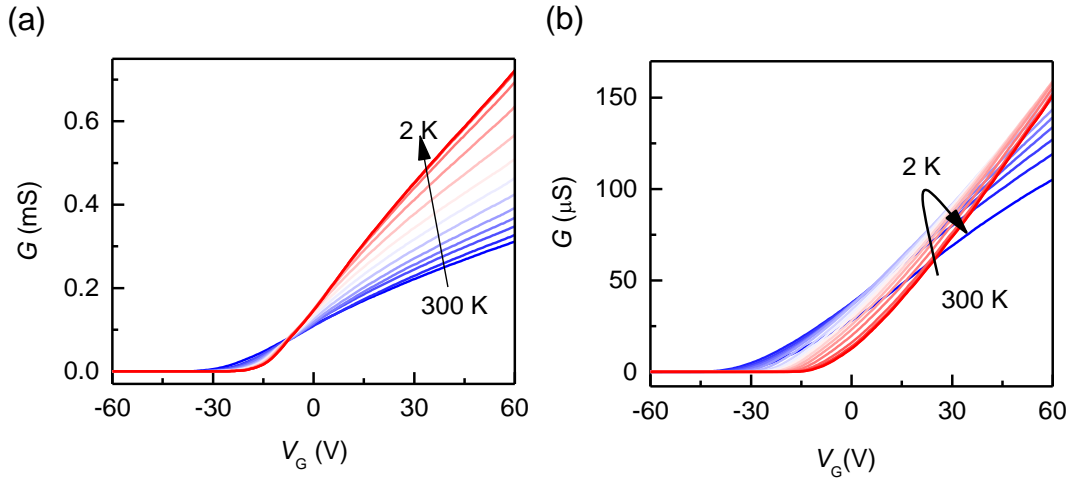

**Figure S2** (a) Conductance as a function of the backgate voltage for an InSe FED with an IIL (sample D). $V_{ds} = 0.1$ V and the channel width and length of sample D are 9 and 10 μm, respectively. (b) Conductance as a function of the gate voltage for a control InSe FED (Sample B in the main text) without an IIL. $V_{ds} = 1$ V and the channel width and length of sample B are 6 and 16 μm, respectively.

Figure S3a compares $\log(I_{ds})$ as a function of $V_G$ for sample A at $T = 2$ K and $T = 250$ K. Sample A exhibits a typical turn-on behavior with a large on/off ratio of $10^9$ and the subthreshold swing of 230 mV/dec at $T = 2$ K. At $T = 250$ K, the on/off ratio and the subthreshold swing of sample A are $10^7$ and 5.6 V/dec, respectively. At elevated temperature, both the off current and the subthreshold swing become larger because of the generation of the thermally excited carriers at off states. On the other hand, the on current decreases due to enhanced phonon scattering at high temperature, resulting in decreased on/off ratio.

Figure S3b shows the $I_{ds}-V_{ds}$ curves of sample C at $V_G = 15$ V with $T$ ranging from 50 to 300 K. All these $I_{ds}-V_{ds}$ curves are linear and symmetric. Because of the small contact barrier enabled by the non-rectifying barrier, it is expected to exhibit even more linear behavior at elevated temperature given that the carriers originated from thermionic emission greatly increases at the contact region, which effectively suppresses the effect of the contact barrier.

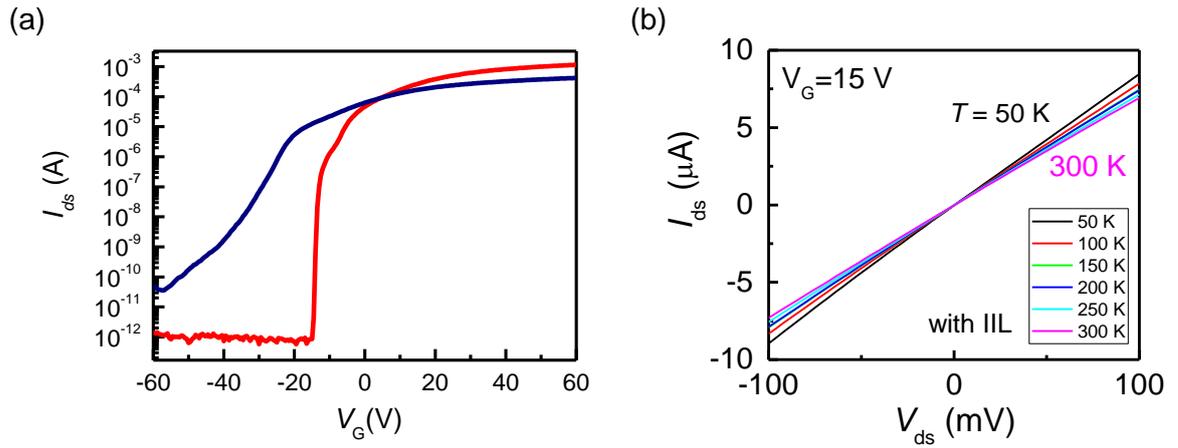

**Figure S3** (a) The logarithmic transfer curve for sample A at $T = 2$ K and $T = 250$ K. $V_{ds} = 1$ V and the channel width and length of sample A are 8 and 13 μm, respectively. (b) The $I_{ds}-V_{ds}$ curves of sample C at $V_G = 15$ V with $T$ ranging from 50 to 300 K. The channel width and length of sample C are 4 and 7 μm, respectively.

We had performed a thickness dependence experiment of thin In layer before the 3 nm thickness was chosen. Here we present the $\mu_{FE}$ as a function of the thickness of In layer for different InSe FEDs in Figure S4. It is found that at the thickness of 3 nm, the InSe FEDs exhibit the highest $\mu_{FE}$. This specific In layer may be related to the optimized reduction effect which was discussed in the XPS study.

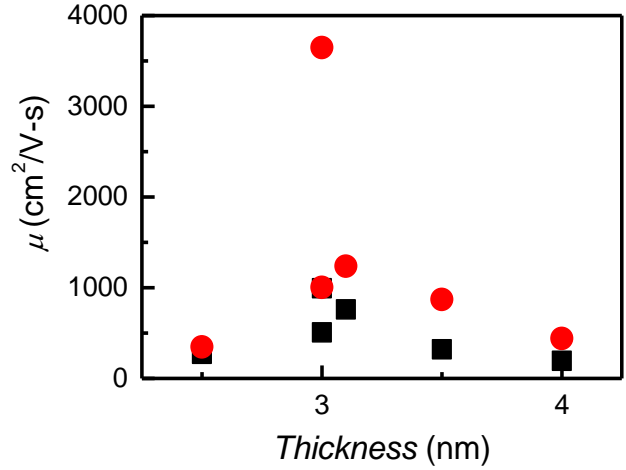

**Figure S4** A distribution of the $\mu_{FE}$ as a function of the thickness of In layer for different InSe FEDs.

In most of the InSe FEDs at high on current regime, the gate leakage current is on the order of 100 pA, which is much smaller than the channel current. For clarity, we show $V_G$ dependence of the drain–source current and the gate leakage current data for a typical InSe sample without IIL in Figure S5a and S5b, respectively, with same $V_G$ range as in Figure 4b. It is noted that the $I_{ds}$ reaches maximum with $V_{ds} = 4$ V and starts to decrease with higher $V_{ds}$, suggesting a heating effect which is not observable in the InSe sample with IIL.

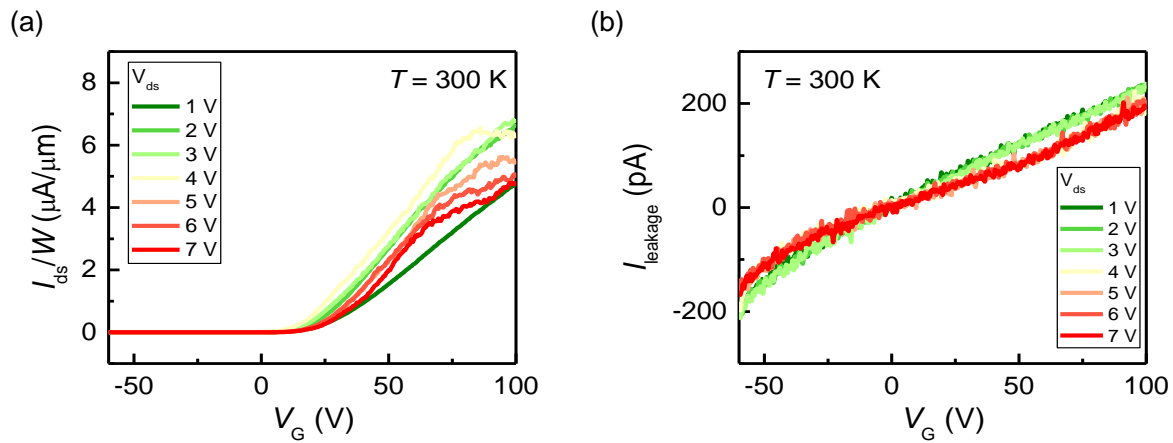

**Figure S5** The $V_G$ dependence of the drain–source current (a) and the gate leakage current (b) data for a typical InSe sample without IIL.

## S3. Estimation of the barrier height in the InSe FEDs

We estimate the effective contact barrier height based on the thermionic emission model.[4] In this model, the drain current injected through the reverse-biased SB includes the thermionic emission current and thermally assisted tunneling current and can be written as (Eq. 1)

$$I_d = A^* T^\alpha \exp\left[-\frac{E_A}{k_B T}\right]\left[1 - \exp(-\frac{qV_{ds}}{k_B T})\right] \tag{1}$$

where $A^*$ is the Richardson constant, $k_B$ is the Boltzmann constant, and α is an exponent of 3/2 for 2D semiconductors. Figure S6a and S6b show the Arrhenius plot of $I_{ds}/T^{3/2}$ for sample A and a control sample without an IIL (Sample B in the main text), respectively. From Figure S6a and S6b, we can plot the activation energy, $E_A$, as a function of $V_G$ for sample A and the control sample (Sample B in the main text), as shown in Figure S6c and S6d (same as Figure 3c and 3d in the main text), respectively. The flat band condition occurs when the band bending vanishes at a critical point, $V_G = V_{FB}$, and at this point, $E_A$ equals the barrier height, $\Phi_B$. Therefore, we can extract values of $\Phi_B = 50$ meV and $\Phi_B = 170$ meV for sample A and the control sample, respectively. The barrier height of the sample with an IIL is smaller, indicating enhanced contact.

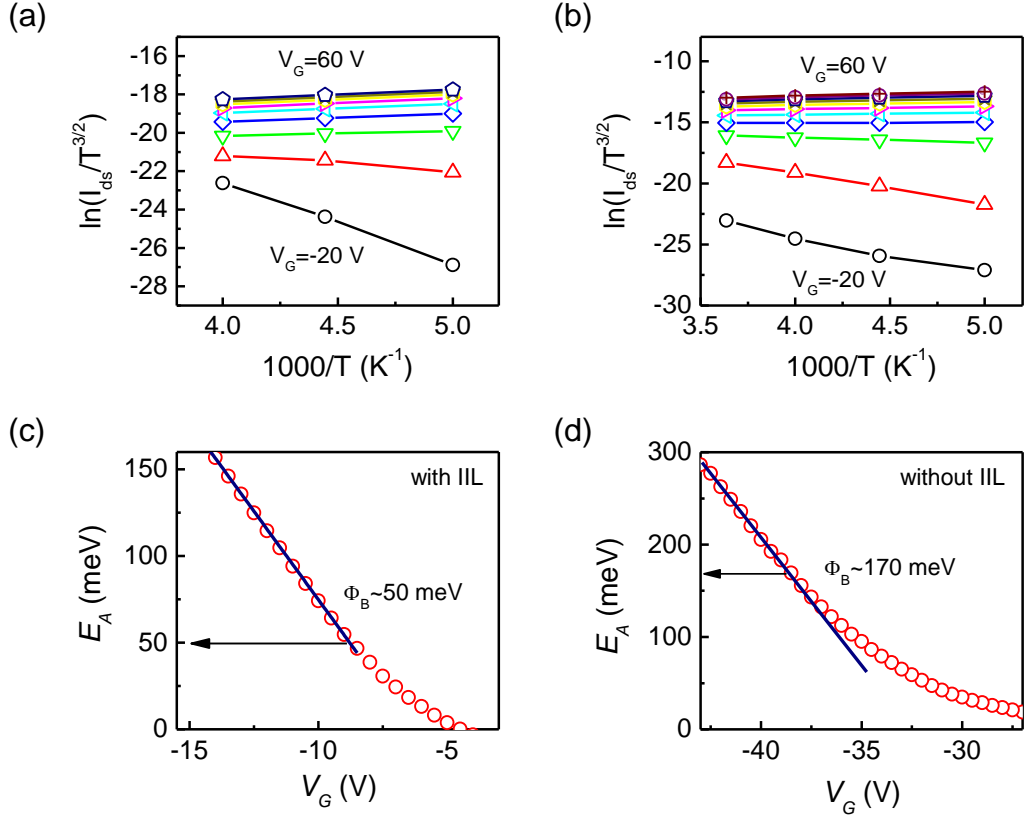

**Figure S6** The Arrhenius plot of $I_{ds}/T^{3/2}$ at different $V_G$ values for (a) sample A and (b) the control sample without an IIL (Sample B in the main text). The $V_G$ dependence of $E_A$ for (c) sample A and (d) the control sample, where the barrier height can be extracted.

To highlight the role of an IIL in the contact enhancement, we show sample statistics for fabricated InSe FEDs with respect to the room $T$ $\mu_{FE}$ and barrier height in Figure S7. The InSe FEDs with an IIL have higher $\mu_{FE}$ values at room $T$ (~680-1040 cm$^2$/V·s) with a smaller contact barrier. In comparison, the InSe FEDs without an IIL exhibit lower $\mu_{FE}$ values at room $T$ (~170-320 cm$^2$/V·s) with a larger contact barrier. This comparison clearly indicates that the IIL can reduce the effective contact barrier, leading to enhanced transport characteristics in two-terminal InSe FEDs.

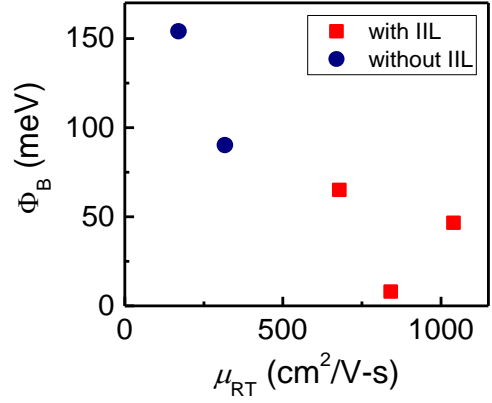

**Figure S7** Distribution of the fabricated InSe FED samples with (red rectangles) and without (blue circles) an IIL with respect to the $\mu_{FE}$ at room $T$ and the barrier height. The applied drain–source voltage is 1 V.

**S4. InSe FED with an IIL deposited only on the channel**

Figure S8a shows a schematic of the control InSe FED with an IIL deposited only on the InSe channel. First, a typical InSe FED is fabricated without an IIL. As shown in Figure S8b, the InSe sample shows a large hysteresis, as described in the inset of Figure 5a in the main text, indicating the presence of trap states at the interfaces.[5] Next, we deposit an IIL only onto the InSe channel, as depicted in the schematic. The InSe sample shows a negligible change in its transport behavior, which is in contrast to the InSe sample with an IIL in the contact region. Moreover, the threshold voltage remains the same before and after the IIL deposition, suggesting no doping effect from the In oxide layer on the InSe channel.

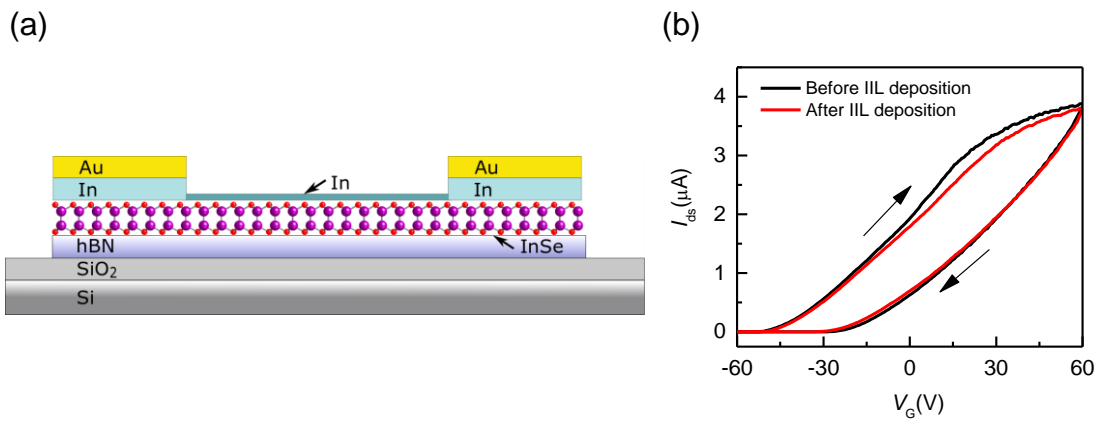

**Figure S8** (a) A schematic of the control InSe FED with an IIL deposited only on the InSe channel. (b) A comparison of the transfer characteristics at room $T$ before and after the deposition of an IIL. The drain–source voltage is 0.1 V.

## S5. $I_{ds}-V_{ds}$ curves of an InSe FED without IIL deposition

Figure S9 shows the $I_{ds}-V_{ds}$ curves at different $V_G$ values for a control InSe FED without an IIL at room $T$. Compared to the InSe FED with an IIL (Figure 4 in the main text), the control InSe sample shows no current saturation behavior and a much smaller breakdown $V_{ds}$. The control InSe sample shows nonlinear $I_{ds}-V_{ds}$ curves, which indicate a larger contact barrier at the metal/InSe interface. Moreover, the drain-source current of the control sample only reaches ~175 µA/µm, which is much smaller than the saturation current of the InSe FED with an IIL.

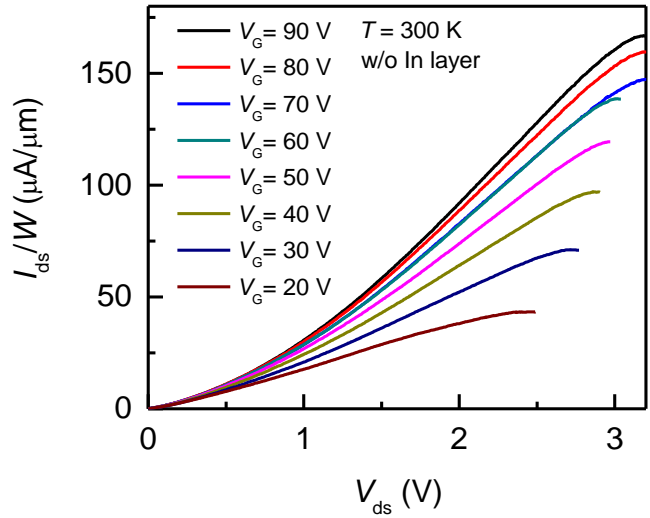

**Figure S9** The output characteristics of a control InSe FED at room $T$ and at different $V_G$ values. The range of the $V_G$ values is from 20 V to 90 V. The channel length of this device is 5 µm, and the channel width is 4.5 µm.

## S6. Estimation of the saturation velocity from the current saturation behavior

We analyze the output characteristics to extract the saturation velocity, $v_{sat}$, by a simple semiempirical model, in which $I_{ds}$ can be expressed as (Eq. 2)

$$I_{ds} = \frac{W}{L}\mu C_{ox}(V_G - V_{th}) \frac{V_{ds}}{[1+(V_{ds}\mu/Lv_{sat})^\alpha]^{1/\alpha}} \quad (2)$$

where $C_{ox} = 11.6$ nF/cm is the gate capacitance and L = 7 μm and W = 3.9 μm are the sample dimensions. Figure S10 shows the calculated and measured $I_{ds}$ values, and a reasonable fitting is attained. $\mu_{FE} = 500$ cm²/V·s at room $T$ and $V_{th} = -24.3$ V are extracted from the transfer characteristics shown in Figure 3b in the main text. We can extract the saturation velocity as $v_{sat} = 1.2 \times 10^7$ cm/s for $V_G = 80$ V. $\alpha = 1.8$ is extracted from the fitting.

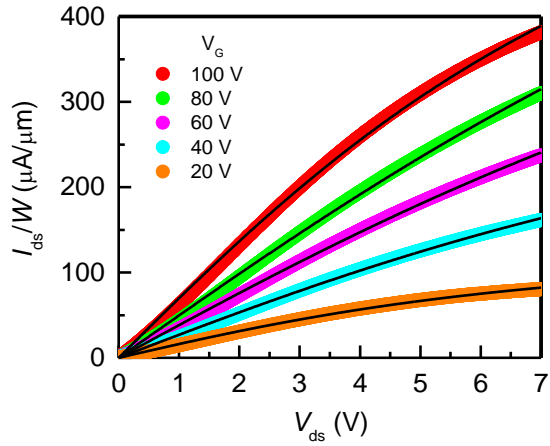

**Figure S10** The output characteristics of an InSe FED (sample A in the main text) at room $T$ with $V_G = 20, 40, 60, 80,$ and $100$ V. The measured $I_{ds}$ values is reasonably described by the model.

## S7. XPS study of InSe oxidation

We characterized the oxidation of InSe flakes by an XPS study, as shown in Figure 5b and 5c. Figure S11a and S11b compare the In 3d and Se 3d core levels before and after a 5-day exposure to ambient conditions. The peaks at 452.8 and 445.2 eV can be attributed to the core levels of In $3d_{3/2}$ and In $3d_{5/2}$, respectively. The peaks at 55.4 and 54.6 eV can be attributed to the core levels of Se $3d_{3/2}$ and Se $3d_{5/2}$, respectively. After the oxidation process, all the characteristic peaks, including In $3d_{3/2}$, In $3d_{5/2}$, Se $3d_{3/2}$, and Se $3d_{5/2}$, exhibit a consistent blueshift of 1.1 eV. The fact that all four core levels change with the same energy suggests a common origin for the oxidation effect.[3]

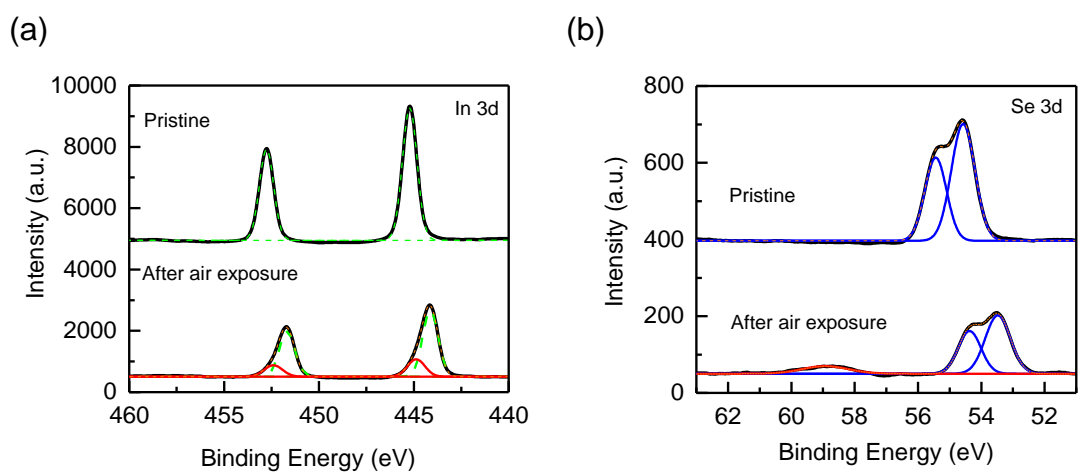

**Figure S11** (a) The XPS data show a comparison of the In 3d peaks before and after exposure to ambient conditions for 5 days. (b) A comparison of the Se 3d peaks before and after exposure to ambient conditions for 5 days.

## S8. Implications of the reduction effect and estimation of the transfer length

To understand the influence of the InSe interfacial oxide reduction on the contact enhancement, we first estimate the lateral conduction range under the metallic contact. The current transfer length, $L_T$, represents the average distance of carrier transport in the contact region, which can be expressed as $L_T = (r_C/\rho^{2D})^{1/2}$, where $r_C$ is the semiconductor/metal interface resistivity and $\rho^{2D}$ is the sheet resistivity.[4] We calculate a large $L_T$ of ~1.25 µm (see next paragraph), which may be attributed to the small $\rho^{2D}$ of the InSe layers, resulting from the intrinsic InSe property and the h-BN substrate.[6-7] Because the carriers transport extensively under the contact region before entering the metallic electrode, the reduction of the top layered InSe plays an important role in reducing the carrier scattering in the contact regime, as depicted in a schematic in Figure S12, leading to the realization of high-performance InSe FEDs with an IIL.

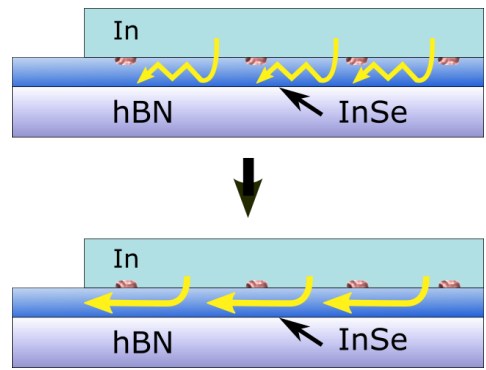

**Figure S12** A schematic of the contact region of an InSe FED showing the effect of the InSe interfacial oxide reduction on the contact enhancement.

To estimate the current transfer length, $L_T$, and the contact resistance, $R_C$, we fabricated a four-terminal InSe FED using a resist-free method.[8] Figure S13a and 13b show the OM images of the InSe FED before and after fabrication of the contact electrodes, respectively. Figure S13c shows the $V_G$-dependent conductance obtained from two- and four-terminal measurements at room $T$. By comparing the two- and four-probe measurements, the contact resistance can be calculated by the equation (Eq. 3)

$$R_c = \frac{1}{2}(R_{2p} - R_{4p}\frac{L}{l}) \qquad (3)$$

where L represents the distance between two outside probes and $l$ represents the distance between two inner probes. Figure S13d shows the $V_G$-dependent contact resistance, where the contact resistance is tuned by $V_G$. Here, we assume that the contact length (> 2 μm) is much larger than $L_T$. Therefore, the contact resistance can be written as $R_c = (\rho^{2D} r_C)^{1/2}$, and the transfer length becomes $L_T = R_C/\rho^{2D}$.[4] By using a sheet resistivity of 6.67 kΩ $\square^{-1}$ and a contact resistance of 10 kΩ μm measured at $V_G = 45$ V, we can estimate the transfer length of $L_T = 1.5$ μm, as described in the previous paragraph.

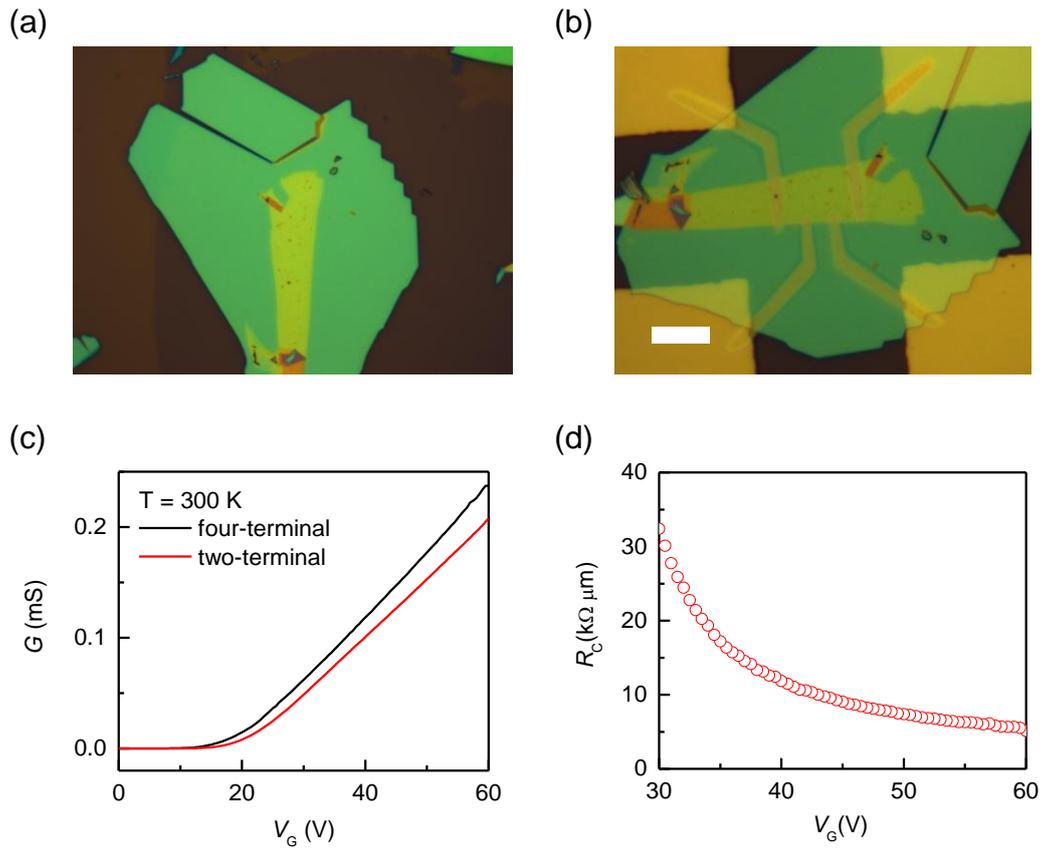

**Figure S13** OM images of InSe/h-BN with an IIL before (a) and after (b) fabrication of the contact electrodes. L= 12 μm, W= 9.6 μm, and $l$= 3 μm. The scale bar is 10 μm. (c) Conductance as a function of the backgate voltage measured with 2-terminal and 4-terminal geometries. (d) Contact resistance as a function of the backgate voltage at room $T$.

**S9. Chemical stability of the InSe samples studied by Raman spectroscopy**

We show that the chemical stability of the InSe FEDs can be improved by the IIL coverage. We performed a stability test under a low vacuum environment ($P{\sim}100$ Torr). Figure S14a compares the transfer characteristics of an InSe FED (sample E) as-fabricated with those of a sample that was kept in a low vacuum for 4 months. The transport characteristics are generally preserved. The MIT is still distinctly observed in sample E after 4 months, and $V_{MIT} - V_{th}$ negligibly changes. $\mu_{FE}$ decreases by 12% and 13% for $T = 2$ K and room $T$ as a result of increased carrier scattering, which is possibly due to surface oxides.

We further present the evolution of the Raman spectra (Figure S14b) to show the chemical stability of the InSe FEDs with an IIL. An InSe sample was partially capped with an IIL, and the OM image is shown in the inset of Figure S14b. The pristine InSe exhibits four characteristic peaks at 115, 178, 199, and 226 cm$^{-1}$, which are attributed to the Raman modes of $A_1'(\Gamma_1^2)$, $E'(\Gamma_3^1)$- TO, $A_2''(\Gamma_1^1)$- LO, and $A_1'(\Gamma_1^3)$, respectively.[1] For uncapped InSe, the $A_1'(\Gamma_1^2)$ and $A_1'(\Gamma_1^3)$ peaks become very small, and the other characteristic peaks vanish after 36 hours in ambient conditions, indicating the degradation of the InSe material. In contrast, all four characteristic peaks are still present after 36 hours in ambient conditions for the InSe FED, indicating that the IIL can modestly protect the InSe surface from oxidation, which is beneficial for the transport characteristics of InSe FEDs.

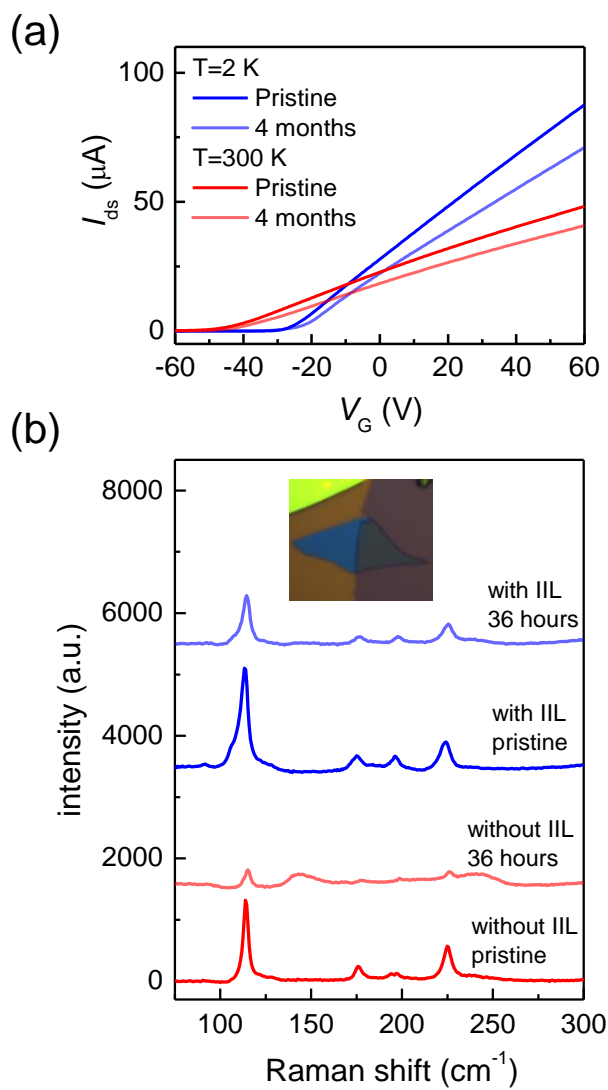

**Figure S14** (a) A comparison of the transfer characteristics of sample E as-fabricated (blue curve, $T = 2$ K; red curve $T = 300$ K), and those of a sample that was kept in a low vacuum for 4 months (light blue curve, $T = 2$ K; light red curve $T = 300$ K). The Raman spectra corresponding to InSe covered with an IIL before (blue curve) and after 36 hours in ambient conditions (light blue curve). (b) A comparison of the Raman spectra corresponding to InSe without an IIL before (red curve) and after 36 hours in ambient conditions (light red curve). Inset: an OM image of an InSe sample partially capped with an IIL.

**References**

(1) Lei, S. D.; Ge, L. H.; Najmaei, S.; George, A.; Kappera, R.; Lou, J.; Chhowalla, M.; Yamaguchi, H.; Gupta, G.; Vajtai, R.; Mohite, A. D.; Ajayan, P. M. Evolution of the Electronic Band Structure and Efficient Photo-Detection in Atomic Layers of Inse. *Acs Nano* **2014,** *8* (2), 1263-1272.

(2) Yang, Z.; Jie, W.; Mak, C.-H.; Lin, S.; Lin, H.; Yang, X.; Yan, F.; Lau, S. P.; Hao, J. Wafer-Scale Synthesis of High-Quality Semiconducting Two-Dimensional Layered Inse with Broadband Photoresponse. *ACS nano* **2017,** *11* (4), 4225-4236.

(3) Ho, P. H.; Chang, Y. R.; Chu, Y. C.; Li, M. K.; Tsai, C. A.; Wang, W. H.; Ho, C. H.; Chen, C. W.; Chiu, P. W. High-Mobility Inse Transistors: The Role of Surface Oxides. *Acs Nano* **2017,** *11* (7), 7362-7370.

(4) Allain, A.; Kang, J. H.; Banerjee, K.; Kis, A. Electrical Contacts to Two-Dimensional Semiconductors. *Nat Mater* **2015,** *14* (12), 1195-1205.

(5) Late, D. J.; Liu, B.; Matte, H. S. S. R.; Dravid, V. P.; Rao, C. N. R. Hysteresis in Single-Layer Mos2 Field Effect Transistors. *Acs Nano* **2012,** *6* (6), 5635-5641.

(6) Bandurin, D. A.; Tyurnina, A. V.; Yu, G. L.; Mishchenko, A.; Zolyomi, V.; Morozov, S. V.; Kumar, R. K.; Gorbachev, R. V.; Kudrynskyi, Z. R.; Pezzini, S.; Kovalyuk, Z. D.; Zeitler, U.; Novoselov, K. S.; Patane, A.; Eaves, L.; Grigorieva, I. V.; Fal'ko, V. I.; Geim, A. K.; Cao, Y. High Electron Mobility, Quantum Hall Effect and Anomalous Optical Response in Atomically Thin Inse. *Nat Nanotechnol* **2017,** *12* (3), 223-227.

(7) Kang, P.; Michaud-Rioux, V.; Kong, X. H.; Yu, G. H.; Guo, H. Calculated Carrier Mobility of H-Bn/Gamma-Inse/H-Bn Van Der Waals Heterostructures. *2d Mater* **2017,** *4* (4).

(8) Shih, F. Y.; Chen, S. Y.; Liu, C. H.; Ho, P. H.; Wu, T. S.; Chen, C. W.; Chen, Y. F.; Wang, W. H. Residue-Free Fabrication of High-Performance Graphene Devices by Patterned Pmma Stencil Mask. *Aip Adv* **2014,** *4* (6), 067129-1-067129-6.